\documentclass[12pt]{article}
\usepackage[table,xcdraw]{xcolor}  
\usepackage[T1]{fontenc}
\usepackage[english]{babel}
\usepackage{lmodern}

\usepackage[sfdefault]{libertinus}  
\usepackage{newtxtext,newtxmath}    

\usepackage{amsmath}
\usepackage{amssymb}
\usepackage{amsthm}

\usepackage{graphicx}
\usepackage{epsfig}
\usepackage{epstopdf}
\usepackage{caption}
\usepackage{subcaption}
\usepackage{smartdiagram}

\usepackage{algorithm}
\usepackage{algpseudocode}

\usepackage{booktabs}
\usepackage{multirow}
\usepackage{longtable}
\usepackage{array}
\usepackage{adjustbox}

\usepackage{setspace}
\usepackage{titlesec}
\usepackage{indentfirst}
\usepackage{enumitem}
\usepackage{float}
\usepackage{here}
\usepackage{lipsum}
\usepackage{geometry}
\usepackage{comment}

\usepackage{hyperref}
\usepackage{cite}
\usepackage{url}

\titleformat{\section}[block]{\rmfamily\bfseries}{\thesection}{1em}{}

\titlespacing*{\section}{0pt}{0pt}{0pt}

\newcommand{\headingfont}{\rmfamily}

\begin{document}

\begin{center}
{\headingfont\fontsize{18}{22}\selectfont\textbf{Non-Heuristic Selection via Hybrid Regularized and Machine Learning Models for Insurance}}
\end{center}

\begin{center}
\headingfont\textbf{Luciano Ribeiro Galvão}$^{1*}$ \\
\headingfont\textbf{Rafael de Andrade Moral}$^2$  \\[2ex]

$^1$ Department of Exact Sciences, University of S\~ao Paulo, Brazil\\
$^2$ Department of Mathematics and Statistics, Maynooth University, Ireland\\
$^*$ corresponding author: lucianogalvao@usp.br\\

\end{center}

\begin{center}
\headingfont\textbf{\textit{Abstract}}
\end{center}

\begin{singlespacing}
In this study, machine learning models were tested to predict whether or not a customer of an insurance company would purchase a travel insurance product. For this purpose, secondary data provided by an open-source website that compiles databases from statistical modeling competitions were used. The dataset used presents approximately 2,700 records from an unidentified company in the tourism insurance sector. Initially, the feature engineering stage was carried out, which were selected through regularized models: Ridge, Lasso and Elastic-Net. In this phase, gains were observed not only in relation to dimensionality, but also in the maintenance of interpretative capacity, through the coefficients obtained. After this process, five classification models were evaluated (Random Forests, XGBoost, H2O GBM, LightGBM and CatBoost) separately and in a hybrid way with the previous regularized models, all these stages using the k-fold stratified cross-validation technique. The evaluations were conducted by traditional metrics, including AUC, precision, recall and F1 score. A very competitive hybrid model was obtained using CatBoost combined with Lasso feature selection, achieving an AUC of 0.861 and an F1 score of 0.808. These findings motivate us to present the effectiveness of using hybrid models as a way to obtain high predictive power and maintain the interpretability of the estimation process.
\end{singlespacing}

\vspace{0,60cm}

\textit{{\headingfont\textbf{Key Words:}} }
\noindent
Convex optimization;
non-linear algorithms;
White Box models;
Model interpretability;
Black Box Models.

\vspace{0,60cm}

\section{\headingfont Introduction}

\begin{spacing}{1.5}
\noindent

Travel insurance is a strategic challenge for the insurance industry, with a significant impact on the customer lifecycle and income generation \cite{cnseg2024}. Understanding the factors that influence the purchase decision and being able to predict customer behavior are essential to develop targeted offers and improve business results. In this context, machine learning (ML) methods have shown promise in modeling complex data sets and improving predictive capabilities in the financial services and insurance sectors \cite{accenture2023}.

On this topic, we aim to demonstrate the use of modern predictive models to ensure predictability in travel insurance contracting, using data made publicly available through an open source modeling competition. The dataset used comprises 2,697 customer records from a company in the tourism insurance sector and includes a mix of demographic, socioeconomic and behavioral attributes resulting in 10 initial attributes.

For this type of product and, more broadly, the insurance market as a whole, there are vast approaches in the literature, from regressive techniques to classification. These seek to explore Machine Learning methods used in predictive studies of travel insurance purchases \cite{lim2023predicting}, \cite{li2023exploring} and \cite{sahai2023insurance}. Another focus is on studies that present high dimensionality related to the sector, proposing that the solutions applied to the market be expanded in order to bring broader and more accurate views \cite{sun2024research}, \cite{tian2023machine}.

Predictive models with complex structures, often called as "black box", tend to have high levels of accuracy, but, on the other hand, offer little transparency. This lack of clarity can be a hindrance, especially in regulatory or strategic contexts or where the results require an understanding of the components of the models. These can be fundamental for taking concrete market actions. The use of five ensemble-based machine learning algorithms is therefore proposed: Random Forest, XGBoost, H2O GBM, LightGBM and CatBoost, associated with regularized models to ensure predictive power and interpretability \cite{breiman2001random,chen2016xgboost,ke2017lightgbm}.

In this work, we explore non-heuristic statistical approaches as an alternative to increase model interpretability. Penalized regression techniques, such as Lasso, Ridge and Elastic Net, available through the glmnet package \cite{glmnet,R} in R, provide a solid basis for variable selection, in addition to efficiently dealing with problems such as multicollinearity. These techniques contribute to simplifying models, reducing dimensionality and highlighting the most relevant predictors. The result of this type of method are models that combine greater interpretability with a more stable and reliable statistical structure \cite{friedman2010regularization,tibshirani1996lasso,zou2005elasticnet}.

Initially, in \ref{methodology}, a general flow of how the construction of hybrid models is demonstrated, going through the definition of regularized models and their evaluation in \ref{regularized} models in isolation and their results can be observed in \ref{results:regularizados}, followed by the blackbox models in \ref{blackbox} whose metrics are shown in \ref{results:blackbox}. In order to compare the results of the individual and combined methodologies, in \ref{results:combined} there is a discussion about these measures, while in \ref{simulation} and \ref{results:simulation} it is demonstrated that the results obtained through simulation are generalizable.

Thus, through this pipeline presented, we attempt to demonstrate the viability of this hybrid approach to reconcile the gains of whitebox and black box models in order to promote new approaches in the insurance sector.\\

\section{Methodology}\label{methodology}

The initial step involved feature engineering, which expanded the original set of 10 attributes to a total of 35 explanatory variables. The original database contains demographic, behavioral, and socioeconomic information such as: the customer's age, annual income, and total miles accumulated, all of which are continuous. There is also the number of family members, which is a discrete covariate. The binary variables were: whether or not the customer had a college degree, chronic illnesses, whether or not they were self-employed, and whether or not they had ever traveled abroad. The target variable was whether or not the customer had purchased travel insurance in the last 24 months.

In addition to the bank's original variables, an expansion was carried out using attribute engineering, with the aim of capturing variations in aspects of each customer's profile. New variables were constructed based on proportions, such as family per capita income and the ratio between income and age, focusing on the individual's socioeconomic level. Scores were also created through weighted combinations of variables, such as the propensity to purchase insurance or not derived from a weighted average between income, history of international travel and flight frequencies associated with empirically applied weights. The perceived risk score, combining factors such as income, flight frequency, normalized age and travel habits also using empirically applied weights, and subsequently normalized $[0.1]$. The family aspect was considered both by per capita income and by creating indicators of low dependency or families with a larger number of members.

In the professional field, variables on job stability, estimated time of experience in the market and specific employment relationships, such as working in the private or public sector, were included. Patterns directly linked to travel were also mapped through the creation of frequency scores, indicators of international experience and interactions between income and participation in loyalty programs of companies in the sector.

Transformations such as normalization and logarithmic scales were applied to variables such as age, income, miles traveled and chronic diseases, with the aim of reducing distortions and stabilizing variances. Finally, aggregated information based on similar groups was added, such as cluster scores generated by grouping algorithms and moving averages by age groups, enriching the modeling with collective references. Details are available in the supplementary materials (\ref{AnexoI}).

\begin{figure}[H]
    \centering
    \includegraphics[width=0.99\linewidth]{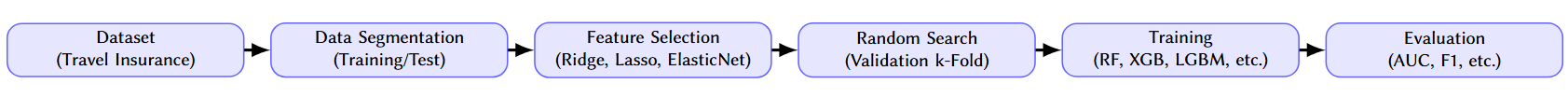} 
    \caption{Flowchart of the data processing, modeling, and validation process}
    \label{fig:fluxo_modelagem}
\end{figure}

After this stage of generating the transformed database, a systematic training and validation process was implemented, using stratified data partitioning via k-fold (5 folds). Following the proposed pipeline, the penalized regression models, Lasso, Ridge and Elastic Net, were applied for feature selection, as illustrated in Figure \ref{fig:fluxo_modelagem}.

Once the feature selection was done, black-box classification models were trained using Random Forest, XGBoost, LightGBM, H2O GBM and CatBoost. These models were chosen due to their ability to handle high-dimensional data and capture linear and non-linear relationships. Hyperparameter optimization was performed via Random Search, a computationally efficient alternative to Grid Search, particularly effective in high-dimensional search spaces \cite{bergstra2012random,bergstra2011algorithms}. This combination is defined in this work as hybrid models, i.e., feature selection via regularization followed by optimization and classification with black-box models.

\subsection{Regularized Models}\label{regularized}

Regularized regression models play a central role in contexts characterized by high dimensionality or the presence of multicollinearity between predictors. In this study, the Ridge, LASSO and Elastic Net methods were applied independently in order to evaluate the best approach. These techniques introduce penalization to the regression coefficients to control the model complexity and reduce the risk of overfitting.

Ridge regression, based on the L2 penalty \cite{hoerl1970ridge}, shrinks coefficients towards zero without making them exactly zero, being particularly effective in scenarios with strong correlation between explanatory variables. LASSO regression (L1 penalty) \cite{tibshirani1996lasso}, in turn, performs variable selection by forcing some coefficients to be exactly zero. 

Elastic Net combines the L1 and L2 penalties through a mixing parameter $\alpha$ \cite{zou2005elasticnet}, offering a hybrid solution. This hybrid approach provides greater flexibility and stability in variable selection, especially in cases involving many correlated features. The pipeline for these models is shown in Algorithm \ref{algor_regula}.

\begin{algorithm}[H]
\caption{Training of Regularized Regression Models with Cross-Validation}
\label{algor_regula}
\begin{algorithmic}[1]
\State \textbf{Input:} Dataset $\mathcal{D} = \{(x_i, y_i)\}_{i=1}^n$, where $x_i \in \mathbb{R}^p$ and $y_i \in \{0,1\}$
\State \textbf{Output:} Sets of selected variables for Ridge, Lasso, and Elastic Net

\State Split $\mathcal{D}$ into stratified training and testing sets: $\mathcal{D}_{\text{train}}$, $\mathcal{D}_{\text{test}}$
\State Set random seed for reproducibility
\State Standardize predictor variables (z-score normalization)

\For{model $\in$ \{\texttt{Ridge}, \texttt{Lasso}, \texttt{Elastic Net}\}}
    \State Set mixing parameter $\alpha$:
    \begin{itemize}
        \item $\alpha = 0$ for Ridge
        \item $\alpha = 1$ for Lasso
        \item $0 < \alpha < 1$ for Elastic Net
    \end{itemize}
    
    \State Apply \texttt{cv.glmnet} with \texttt{family = "binomial"} from the glmnet package \cite{glmnet} for R software:
    \begin{itemize}
        \item Use \texttt{type.measure = "auc"} to select the optimal $\lambda$
        \item Employ stratified $k$-fold cross-validation (e.g., $k=10$)
    \end{itemize}

    \State Extract coefficients $\hat{\beta}$ from the best model (maximum AUC)
    \State Select variables where $\hat{\beta}_j \neq 0$
\EndFor

\State \textbf{Return:} Lists of selected variables for each regularization method

\end{algorithmic}
\end{algorithm}

\subsection{Integration of Regularized Models and 
Non-Linear Algorithms}\label{blackbox}

For each of the regularized approaches, models were trained using random samples
of the dataset, with control over the random seed to ensure the reproducibility of the
results. This step was accompanied by stratified cross-validation to guarantee that the
class distribution was preserved across training and validation folds \cite{kohavi1995crossvalidation}. The tuning process for the hyperparameters $\lambda$ (penalty strength) and $\alpha$ (in the case of Elastic Net) was
carried out based on the maximization of the AUC metric (Area Under the ROC Curve) \cite{bradley1997auc, friedman2010glmnet,hastie2009elements, james2013introduction}.

After selecting the optimal hyperparameters for the regularized models, an analysis
of the estimated coefficients was performed to assess the relative importance of each
variable. This analysis enabled an interpretable understanding of the predictors’ effects
within the travel insurance context, revealing consistent patterns among demographic
and behavioral attributes associated with a higher likelihood of product adoption \cite{tibshirani1996lasso, zou2005elasticnet,chen2016xgboost, dorogush2018catboost, ke2017lightgbm, breiman2001random}.

Based on the feature sets selected by the penalized regression models, a second mod-
eling phase was carried out using algorithms with greater predictive capacity. At this stage, black-box models were explored, including Random Forest, Gradient
Boosting Machines (GBM), XGBoost, CatBoost, and LightGBM  widely recognized for
their robustness and superior performance in structured data scenarios \cite{breiman2001random, chen2016xgboost, ke2017lightgbm, prokhorenkova2018catboost}.

These models were tuned through a random search procedure, aiming to identify the best combination of hyperparameters, including the number of trees, maximum tree depth, learning rate and internal regularizations, such as $l2\_leaf\_reg$ in CatBoost that incorporates native regularization mechanisms such as the \cite{prokhorenkova2018catboost} parameter. This step followed a rigorous experimental protocol, maintaining the previously defined training-testing split and applying cross-validation within the training set for each hyperparameter configuration.

For each black-box model, performance metrics such as AUC, F1-score, precision, and
recall were collected to enable a direct comparison of the gains obtained relative to the
regularized models. Additionally, variable importance analyses were carried out for the
tree-based models using each package’s native methods  including information gain
and permutation-based importance  to assess the consistency of key predictors across
methods \cite{altmann2010permutation, lundberg2017shap}. The entire modeling pipeline is summarized in Algorithm \ref{algor_blackbox}.

\begin{algorithm}[H]
\caption{Training of Black-Box Models with Features Selected via Regularization}
\label{algor_blackbox}
\begin{algorithmic}[1]
\State \textbf{Input:} Dataset $\mathcal{D} = \{(x_i, y_i)\}_{i=1}^n$, where $x_i \in \mathbb{R}^p$ and $y_i \in \{0,1\}$
\State \textbf{Preprocessing:} Split $\mathcal{D}$ into training set ($\mathcal{D}_{\text{train}}$) and test set ($\mathcal{D}_{\text{test}}$), using a fixed random seed

\State \textbf{Step 1: Feature Selection via Regularization}
\State Fit regularized regression models (Ridge, Lasso, or Elastic Net) on $\mathcal{D}_{\text{train}}$ with cross-validation to determine $\lambda^*$
\State Select feature subset $\mathcal{S} = \{x_j : \hat{\beta}_j \ne 0\}$

\State \textbf{Step 2: Training Black-Box Models Using $\mathcal{S}$}
\For{\textbf{model} $\in$ \{\textit{Random Forest}, \textit{XGBoost}, \textit{CatBoost}, \textit{H2O GBM}, \textit{LightGBM}\}}

    \State Define hyperparameter search space and draw:
    \begin{itemize}
        \item Number of trees: $T \sim \text{Uniform}(100, 1000)$
        \item Maximum depth: $d \sim \text{Uniform}(3, 15)$
        \item Learning rate: $\eta \sim \text{LogUniform}(0.001, 0.2)$
        \item Model-specific parameters (e.g., \texttt{l2\_leaf\_reg} for CatBoost)
    \end{itemize}

    \State \textbf{Random Search with $k$-fold Cross-Validation:}
    \For{each randomly sampled 5 times configuration $h$}
        \State Evaluate mean AUC across validation folds on $\mathcal{D}_{\text{train}}$
    \EndFor
    \State Select optimal hyperparameters $h^*$

    \State Train final model using $h^*$ on $\mathcal{D}_{\text{train}}$
    \State Evaluate test performance on $\mathcal{D}_{\text{test}}$: AUC, F1-score, Precision, Recall
    \State Estimate feature importance (e.g., information gain, Gini index, permutation)

\EndFor

\State \textbf{Output:} Final trained models using $\mathcal{S}$, optimal hyperparameters $h^*$, test metrics, and variable importance rankings

\end{algorithmic}
\end{algorithm}

The complete pipeline of this analysis algorithm, including all scripts used in this study, are available in the GitHub repository \cite{galvao2025hybridmodels}.\\

\section{Results}\label{resultados}
Subsection \ref{results:regularizados} presents the results obtained from the application of penalized regression models Ridge, Lasso, and Elastic Net in the task of predicting a binary response variable. These models were selected for their ability to perform variable selection or shrinkage through coefficient penalization, which helps to mitigate overfitting while also identifying the covariates with the greatest contribution to the response.  
In the visual presentation, it was decided to make a cut of the 10 most relevant variables considering the range of coefficients, seen in \ref{fig:comparativo_coeficientes}, while also presenting comments on the interpretability of the coefficients.

\subsection{Regularized Models}\label{results:regularizados}

The analysis of the regularized models was performed on three main axes: the relative importance of the variables, the trajectory of the AUC metric as a function of the penalty parameter $\lambda$ (via cross-validation) and the confusion matrices derived from the performance on the test set.

\begin{figure}[H]
\centering

\begin{subfigure}[b]{0.40\textwidth}
    \centering
    \includegraphics[width=\textwidth]{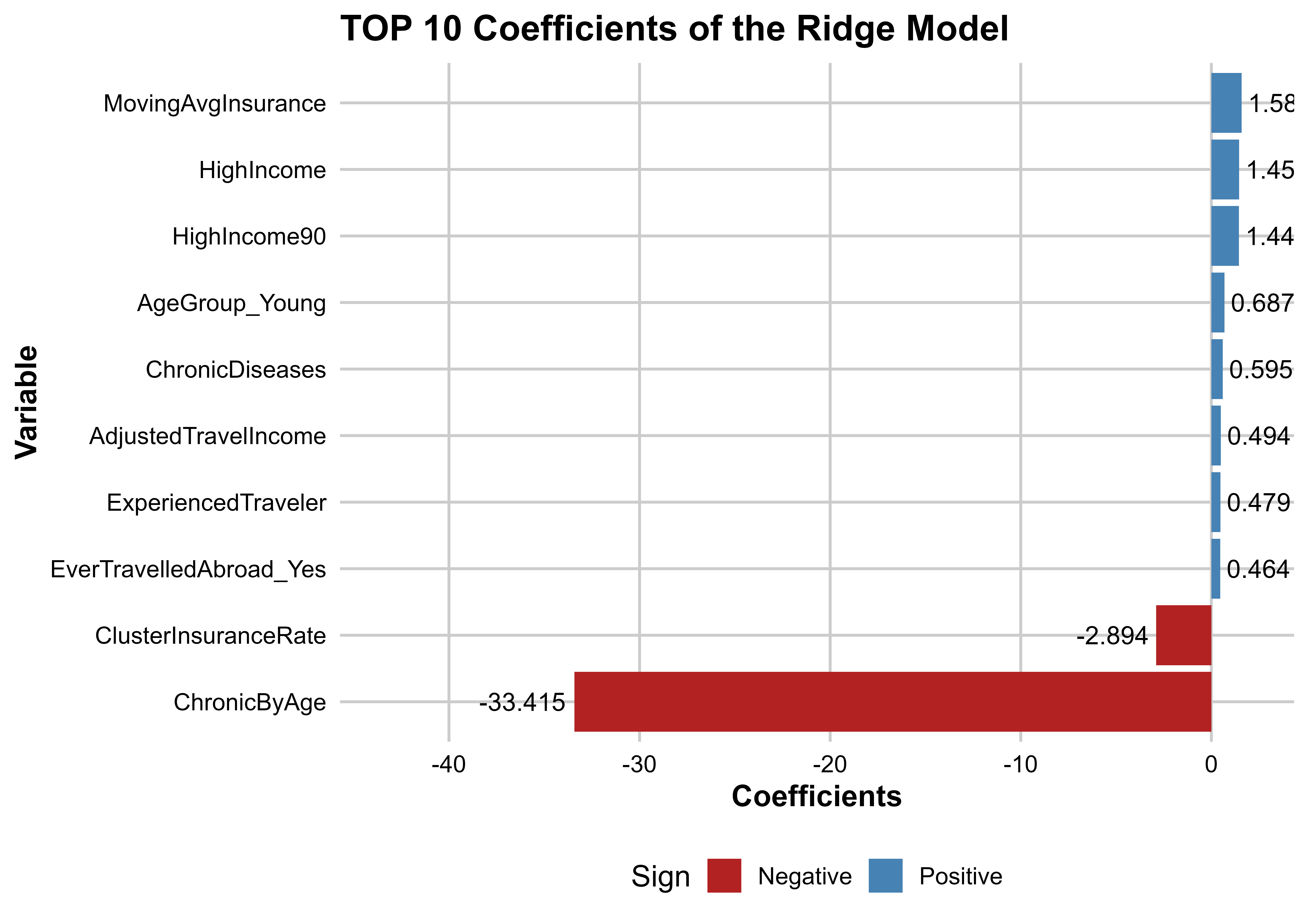}
    \subcaption{Ridge}
    \label{fig:top10_ridge}
\end{subfigure}
\hfill
\begin{subfigure}[b]{0.4\textwidth}
    \centering
    \includegraphics[width=\textwidth]{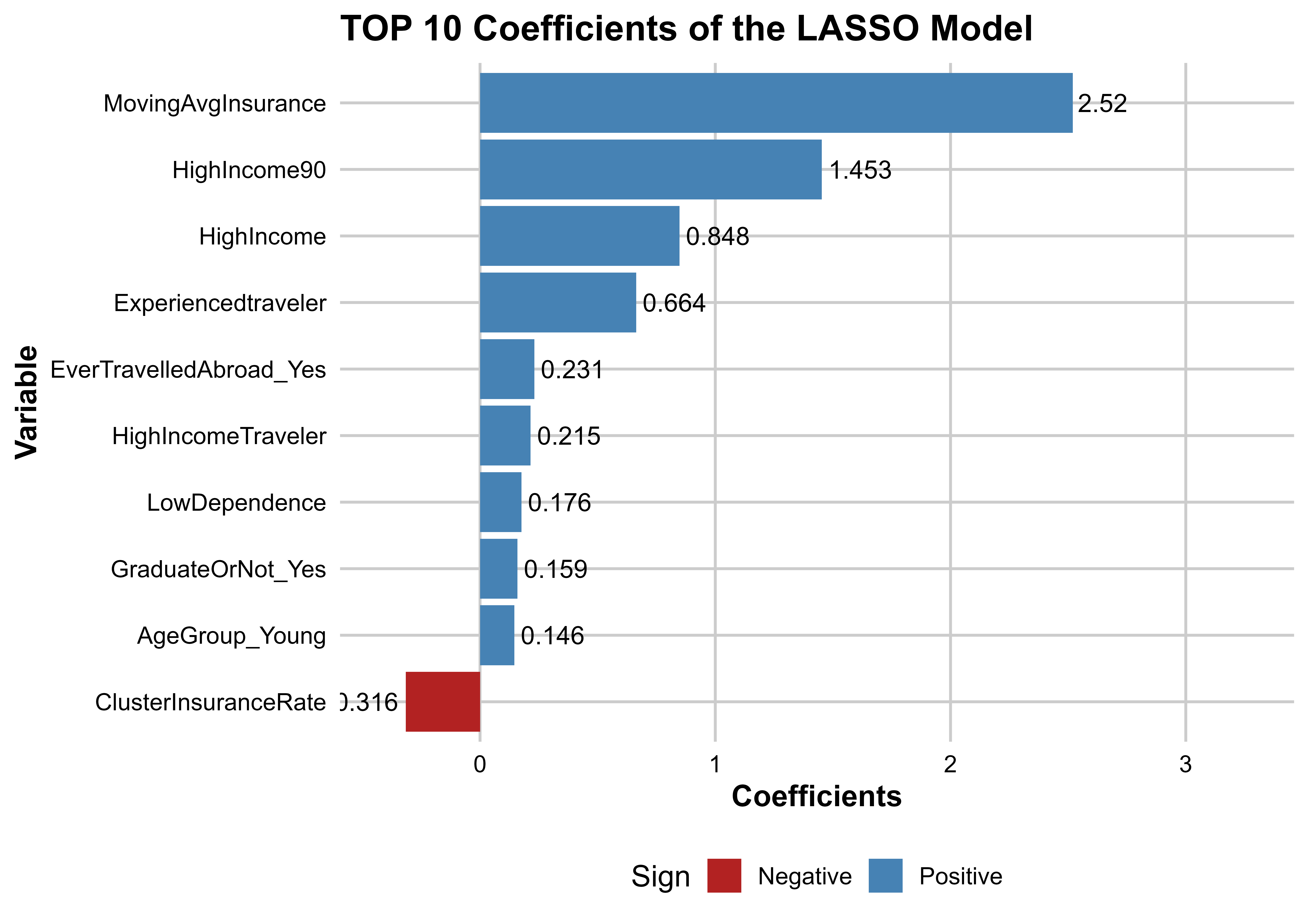}
    \subcaption{Lasso}
    \label{fig:top10_lasso}
\end{subfigure}
\vspace{0.5cm}

\begin{subfigure}[b]{0.40\textwidth}
    \centering
    \includegraphics[width=\textwidth]{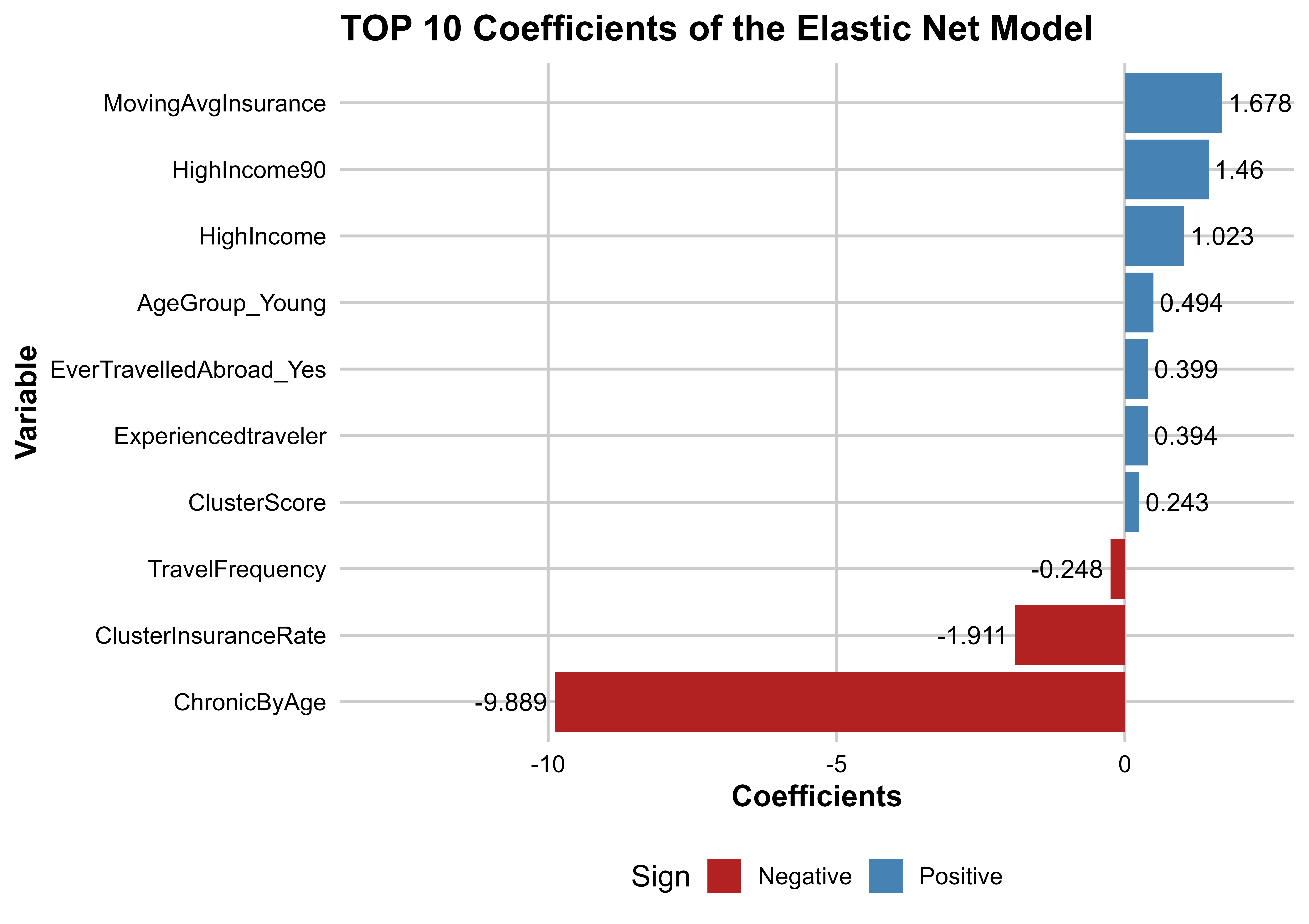}
    \subcaption{Elastic Net}
    \label{fig:top10_elasticnet}
\end{subfigure}

\caption{Top 10 most relevant variables based on the magnitude of the estimated coefficients in each regularized model based on AUC increment.}
\label{fig:comparativo_coeficientes}
\end{figure}

Figure \ref{fig:comparativo_coeficientes} shows the top 10 coefficients estimated by the methods. The regularized Ridge model \ref{fig:top10_ridge} assigns non-zero weights to all highlighted variables, reflecting their continuously decreasing nature, while the Lasso model \ref{fig:top10_lasso} promotes greater parsimony by acting by nullifying coefficients; finally, the Elastic Net model \ref{fig:top10_elasticnet} has the characteristic of combining both strategies, retaining relevant variables even in the presence of multicollinearity.

The coefficients estimated by the regularized models demonstrated good performance in identifying relevant variables. We highlight the variable ChronicByAge, which was demonstrated to be important in two of the three models. In the Ridge model, this variable presented the highest coefficient, considering absolute values, (-33.4), indicating a strong negative association with the basic variable. L2 regularization showed that the explanatory variable ChronicByAge is relevant. The value of this coefficient suggests that individuals with a greater accumulation of age-related comorbidities are less likely to purchase insurance.

The results observed for the Lasso model show that ChronicByAge does not appear among the top 10 predictors. This indicates that the L1 penalty imposed a stricter filter when compared to Ridge, highlighting variables with more isolated effects and less collinearity. In the Elastic Net model, which combines L1 and L2 regularizations, the variable chronic diseases weighted by age reappears with a substantial effect (-9.889). It can then be inferred that its absence in the Lasso model may be due to strong multicollinearity with other variables in the model. In addition, variables such as MovingAgInsurance,HighIncome ExperiencedTraveler and AgeGroup stand out among the main positive coefficients in the three models, showing a directly proportional influence. These variables are associated with a greater probability of purchasing travel insurance, reinforcing the role of income, previous travel experience and age as influencers in the profile of travelers inclined to purchase insurance.

\begin{figure}
    \centering
    \includegraphics[width=0.9\linewidth]{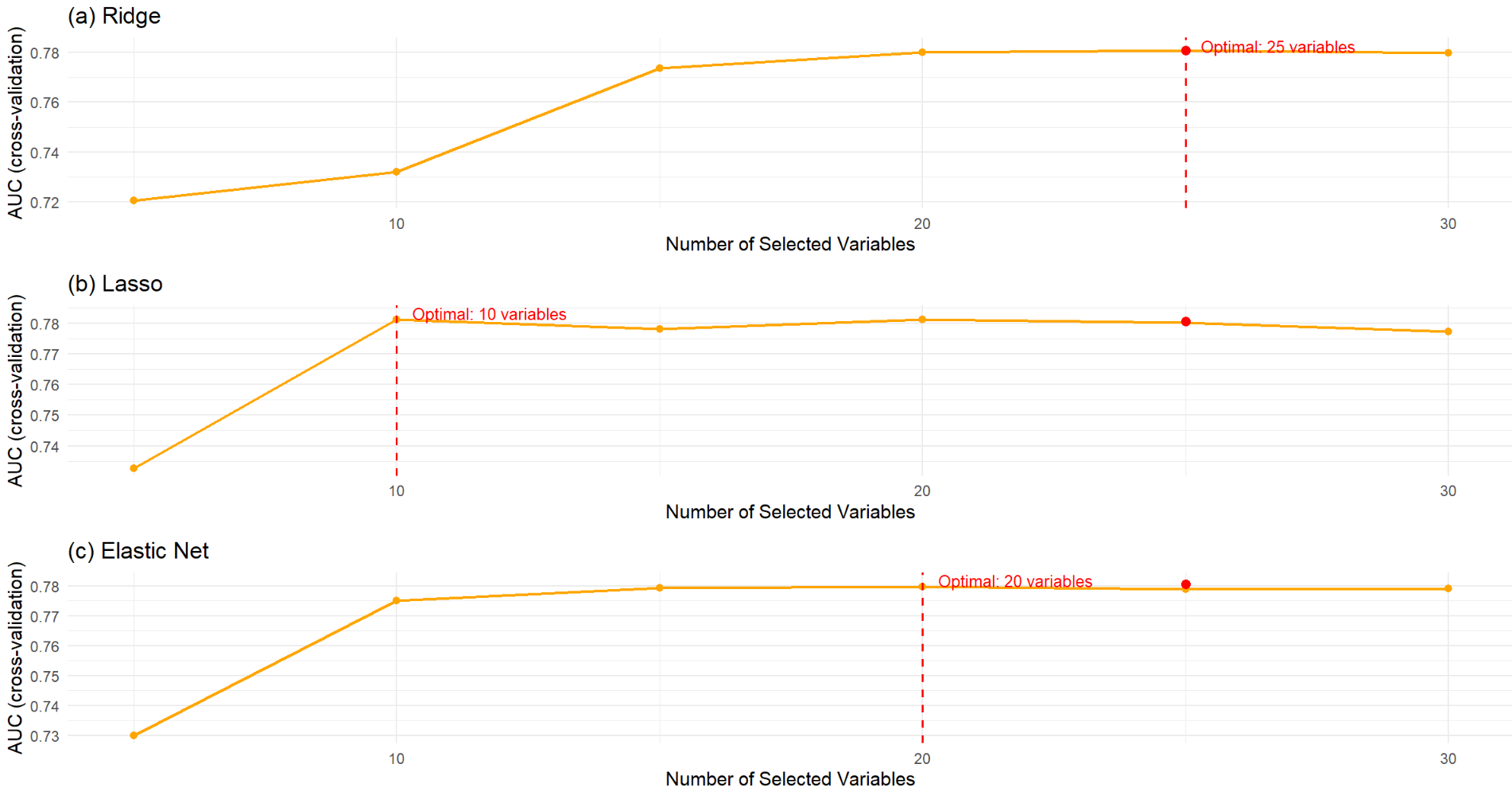}
    \caption{AUC performance curves as a function of the number of selected variables for the Ridge, Lasso, and Elastic Net models. The results were obtained using stratified cross-validation.}
    \label{fig:comparativo_auc}
\end{figure}

Next, Figure \ref{fig:comparativo_auc} shows the evolution of the AUC value as a function of the number of selected variables, based on the importance criteria derived from the regularized models. Evaluating the Ridge model (a), an improvement in performance is observed until reaching stabilization with a larger set of variables compared to the others. The Lasso model (b), reaches a high AUC value in a few cycles considering the number of variables, while (c) presents an intermediate curve, due to its balance between the L1 and L2 penalties, as more variables are included.

In order to evaluate the metrics of the regularized models, the table \ref{tab:confusion_matrices_classic} presents the confusion matrices of the three regularized models when using a training/testing split (80:20), highlighting the correct and incorrect classifications in the binary prediction task of the test set.

All models showed similar sensitivity, demonstrating equivalent capacity in correctly identifying the positive cases tested. The Lasso model demonstrated to have the highest specificity and precision among those evaluated, having a more accurate classification factor for this data set, as seen in the table \ref{tab:evaluation_metrics} with an accuracy of 0.757. Ridge showed lower precision, despite a similar AUC, while the Elastic Net maintained intermediate performance as a hybrid solution between parsimony and robustness.

\begin{table}[H]
\centering
\caption{Confusion matrices for the Ridge, Lasso, and Elastic Net models with True Positive (TP), False Negative (FN), False Positive (FP), and True Negative (TN) numbers.}
\label{tab:confusion_matrices_classic}
\resizebox{\textwidth}{!}{%
\begin{tabular}{l|cc|cc|cc}
\toprule
\multirow{2}{*}{} & \multicolumn{2}{c|}{\textbf{Ridge}} & \multicolumn{2}{c|}{\textbf{Lasso}} & \multicolumn{2}{c}{\textbf{Elastic Net}} \\
\cmidrule(lr){2-3} \cmidrule(lr){4-5} \cmidrule(lr){6-7}
 & \textbf{Positive (S)} & \textbf{Negative (N)} & \textbf{Positive (S)} & \textbf{Negative (N)} & \textbf{Positive (S)} & \textbf{Negative (N)} \\
\midrule
\textbf{Predicted Positive (S)} & 187 (TP) & 97 (FN)  & 187 (TP) & 97 (FN)  & 187 (TP) & 97 (FN) \\
\textbf{Predicted Negative (N)} & 39 (FP)  & 216 (TN) & 34 (FP)  & 221 (TN) & 37 (FP)  & 218 (TN) \\
\bottomrule
\end{tabular}%
}
\end{table}

The \ref{tab:evaluation_metrics} table summarizes the overall performance metrics, including balanced precision, which is used to mitigate cases of class imbalance because it averages sensitivity and specificity, unlike standard precision, which can bias towards majority classes. The results corroborate previous observations: the Ridge model achieved the highest \textit{recall} and AUC, the Lasso model had the best \textit{precision}, and the Elastic Net achieved intermediate performance. These results demonstrate the impact of the choice of the regulation model on several indicators relevant to classification models that use transformation.

\begin{table}[H]
\centering
\caption{Evaluation metrics for the Ridge, Lasso, and Elastic Net models on the test set.}
\label{tab:evaluation_metrics}
\resizebox{\textwidth}{!}{%
\begin{tabular}{lccccccc}
\toprule
\textbf{Model} &\textbf{AUC} & \textbf{Accuracy} & \textbf{Sensitivity (Recall)} & \textbf{Specificity} & \textbf{Precision (PPV)} & \textbf{F1-score} & \textbf{Balanced Accuracy} \\
\midrule
\textbf{Ridge}      & 0.809 & 0.7477 & 0.6585 & 0.8471 & 0.8274 & 0.7330 & 0.7528 \\
\textbf{Lasso}      & 0.807 & 0.7570 & 0.6585 & 0.8667 & 0.8462 & 0.7402 & 0.7626 \\
\textbf{Elastic Net} & 0.809 & 0.7514 & 0.6585 & 0.8549 & 0.8348 & 0.7362 & 0.7567 \\
\bottomrule
\end{tabular}%
}
\end{table}

\subsection{Black-Box Models}\label{results:blackbox}

In this section we will present a comparison where all the steps dimensionality reduction, selection of best features and hyperparameter optimization methods were performed using only black box models, that is, without considering regularization techniques here in order to have a comparative means between joint and isolated methods.

\begin{table}[H]
\centering
\caption{Performance and Optimized Hyperparameters of Black-Box Models in Travel Insurance Prediction}
\label{tab:modelosblackbox}
\resizebox{\textwidth}{!}{%
\begin{tabular}{lrrrrr|l}
\toprule
\textbf{Algorithm} & \textbf{\ Vars} & \textbf{AUC} & \textbf{Precision} & \textbf{Recall} & \textbf{F1 Score} & \textbf{Hyperparameters} \\
\midrule
Random Forest & 12 & 0.8874 & 0.8022 & 0.8431 & 0.8222 & iterations = 185, mtry = 10 \\
XGBoost       & 8  & 0.8978 & 0.7788 & 0.9529 & 0.8571 & max\_depth = 3, eta = 0.0526, nrounds = 239 \\
H2O GBM       & 10 & 0.8985 & 0.8135 & 0.8039 & 0.8087 & max\_depth = 3, ntrees = 310 \\
LightGBM      & 13 & 0.9023 & 0.8097 & 0.8510 & 0.8298 & learning\_rate = 0.0246, nrounds = 256, num\_leaves = 64 \\
CatBoost      & 11 & 0.8986 & 0.7799 & 0.9451 & 0.8546 & learning\_rate = 0.0240, depth = 6, iterations = 403 \\
\bottomrule
\end{tabular}%
}
\end{table}

Table \ref{tab:modelosblackbox} presents the results of the black-box models in the test set. The LightGBM model obtained the best performance compared to the others (AUC = 0.902), also with good indices in precision (0.81) and recall (0.85), resulting in an F1 Score (0.83). Next, we have CatBoost, which also appeared competitive (AUC = 0.899, F1 Score = 0.85 and sensitivity = 0.95). The XGBoost model had the second highest F1 Score (0.86) with a high recall (0.95), but a low precision (0.78) compared to the others, which suggests a tendency towards a higher incidence of false positives. Random Forest presented consistent recall (0.84), but had lower AUC (0.887) and precision (0.80), slightly compromising its F1 Score (0.82). When evaluating the H2O GBM model, a lower performance is observed compared to the others (AUC = 0.899, F1 Score = 0.81 and recall = 0.80), but still with good metrics. These results suggest that models such as LightGBM and CatBoost have better adherence to the data set compared to the others.\\

\section{Combined Approach}\label{results:combined}

This section presents the comparative results between regularized models (Ridge, LASSO and Elastic Net) and black-box models (Random Forest, XGBoost, H2O GBM, LightGBM and CatBoost) applied to the task of predicting whether or not to purchase travel insurance. The objective is to analyze how different modeling strategies, with different levels of interpretability, affect the identification of relevant variables and the predictive performance of the final model. In the regularized models, variable selection was determined by L1 and/or L2 penalties, while in the black-box models, the importance of variables was determined based on internal metrics, such as information gain and noise reduction that did not generate performance gains. For both approaches, the most relevant variable subsets were collected, and hyperparameter tuning was performed through random search with 5-fold stratified cross-validation, each subset was reconditioned in the pipeline aiming at maximizing the AUC. Table \ref{tab:bb_models_reg} summarizes the main results of this analysis.

\begin{table}[H]
\centering
\caption{Performance of Black-Box Models with Feature Selection via Regularization}
\label{tab:bb_models_reg}
\adjustbox{max width=\textwidth}{
\begin{tabular}{l l c c c c l}
\toprule
\textbf{Algorithm} & \textbf{Regularization} & \textbf{AUC} & \textbf{Precision} & \textbf{Recall} & \textbf{F1-Score} & \textbf{Hiperparameters} \\
\midrule
CatBoost     & ElasticNet & 0.8523 & 0.7412 & \textbf{0.8635} & 0.7975 & learning\_rate = 0.2043, depth = 4, iterations = 63 \\
CatBoost     & Lasso      & \textbf{0.8611} & \textbf{0.8024} & 0.8141 & \textbf{0.8082} & learning\_rate = 0.0242, depth = 4, iterations = 389 \\
CatBoost     & Ridge      & 0.8598 & 0.7568 & 0.8177 & 0.7859 & learning\_rate = 0.1666, depth = 5, iterations = 124 \\
H2O GBM      & ElasticNet & 0.8609 & 0.7653 & 0.8100 & 0.7871 & num\_leaves = 4, learning\_rate = 139 \\
H2O GBM      & Lasso      & 0.8565 & 0.7693 & 0.7439 & 0.7562 & num\_leaves = 4, learning\_rate = 167 \\
H2O GBM      & Ridge      & 0.8565 & 0.7633 & 0.7889 & 0.7757 & num\_leaves = 4, learning\_rate = 139 \\
LightGBM     & ElasticNet & 0.8598 & 0.7552 & 0.8135 & 0.7832 & learning\_rate = 0.0196, max\_depth = 337, eta = 52 \\
LightGBM     & Lasso      & 0.8588 & 0.7623 & 0.8058 & 0.7834 & learning\_rate = 0.1130, max\_depth = 63, eta = 64 \\
LightGBM     & Ridge      & 0.8543 & 0.7523 & 0.7096 & 0.7301 & learning\_rate = 0.0780, max\_depth = 78, eta = 52 \\
Random Forest& ElasticNet & 0.8484 & 0.7353 & 0.8415 & 0.7831 & iterations = 141, mtry = 3 \\
Random Forest& Lasso      & 0.8503 & 0.7327 & \textbf{0.8442} & 0.7842 & iterations = 283, mtry = 6 \\
Random Forest& Ridge      & 0.8513 & 0.7380 & 0.8337 & 0.7833 & iterations = 393, mtry = 7 \\
XGBoost      & ElasticNet & 0.8545 & 0.7503 & 0.7771 & 0.7634 & max\_depth = 394, num\_leaves = 3, learning\_rate = 0.0565 \\
XGBoost      & Lasso      & 0.8583 & 0.7531 & 0.7917 & 0.7719 & max\_depth = 214, num\_leaves = 5, learning\_rate = 0.0721 \\
XGBoost      & Ridge      & 0.8584 & 0.7464 & 0.8535 & 0.7961 & max\_depth = 399, num\_leaves = 3, learning\_rate = 0.0317 \\
\bottomrule
\end{tabular}
}
\end{table}

Table \ref{tab:bb_models_reg} compares the performance of black-box models combined with different variable selection strategies via regularization (Elastic Net, LASSO, and Ridge). The combination of CatBoost with Lasso obtained the highest AUC (0.861) and the highest F1 Score (0.80) compared to the other models, with good indices for precision (0.8024) and recall (0.8141). This association suggests in our findings that the power of omitting irreversible features of the regularized LASSO model and the predictive power of the CatBoost classifier model are comparable.

Other hybrid modeling methods were also evaluated, including H2O GBM with Elastic Net, which obtained an AUC of 0.8609 and F1 Score of 0.7871, highlighting the highest recall (0.8183) among all the combination modeling methods listed. Following the table analysis, the XGBoost model combined with Ridge demonstrated a robust overall performance with an AUC of 0.8584 and an F1 score of 0.7961, demonstrating the effectiveness of Ridge even in highly nonlinear scenarios.

Meanwhile, LightGBM and Random Forest performed more moderately. LightGBM demonstrated consistency across regularizations, especially with Lasso (AUC = 0.8588, F1 = 0.7834), while Random Forest excelled mainly in its recovery (Elastic Net: 0.8415; Lasso: 0.8442), but presented comparatively lower AUC scores.

\begin{table}[H]
\centering
\caption{Performance metrics and optimized hyperparameters of the CatBoost model trained using features selected via LASSO regularization. Model evaluated using AUC, precision, recall and F1-Score}
\label{tab:best_model_catboost_lasso}
\begin{tabular}{l c}
\toprule
\textbf{Metric} & \textbf{Value} \\
\midrule
Algorithm & CatBoost \\
Regularization & LASSO \\
AUC & 0.8611 \\
Precision & 0.8024 \\
Recall & 0.8141 \\
F1-Score & 0.8082 \\
\midrule
\textbf{Hyperparameters} & \\
learning\_rate & 0.0242 \\
depth & 4 \\
iterations & 389 \\
\bottomrule
\end{tabular}
\end{table}

Table \ref{tab:best_model_catboost_lasso} presents the performance of the CatBoost model with variable selection via Lasso regularization. This model had the highest AUC (0.861) among all combinations tested here, indicating relevant class discrimination. Of note are the hyperparameters obtained via optimization (learning rate = 0.0242, depth = 4, and 389 booster iterations). These results reinforce the quality of the association between CatBoost's learning capabilities and Lasso's feature selection.

\vspace{0.60cm}

\section{Simulation Studies } \label{simulation}
\subsection{Modeling and Experimental Evaluation}

\begin{figure}[H]
    \centering
    \includegraphics[width=0.3\linewidth]{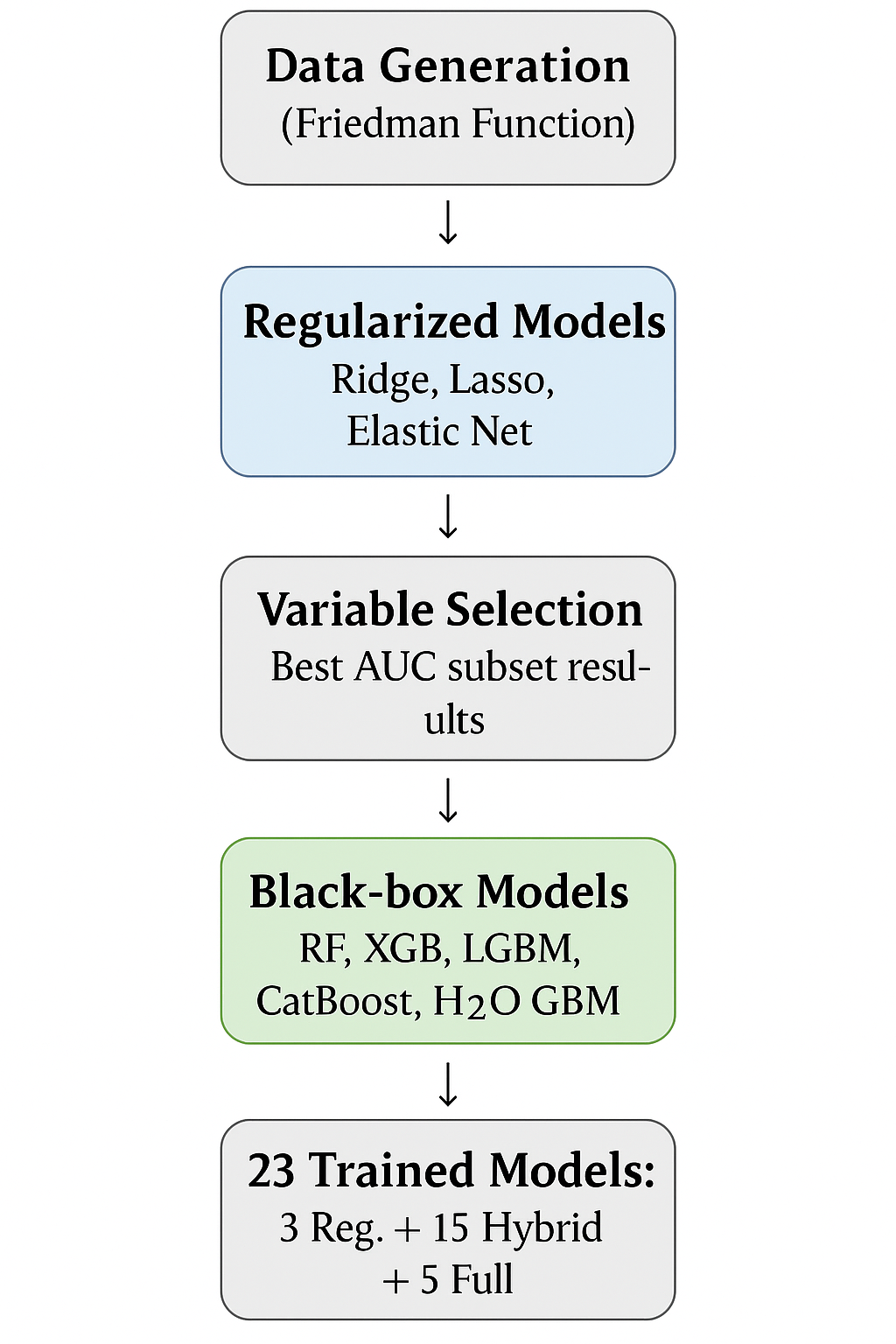}
    \caption{Flowchart for building hybrid models with variable selection via regularization}
    \label{fig:simul_diagram}
\end{figure}

In this study, we evaluate the performance of 23 associated regression models, including three regularized linear models and five nonlinear algorithms, trained on the complete set of variables. We also present in this group of 23, fifteen hybrid models that combine variable selection with nonlinear algorithms again applied to the complete dataset. The models were trained using the full set of covariates generated from the function proposed by \cite{friedman1991}, expressed as:

\begin{equation}
y = 10 \sin(\pi x_1 x_2) + 20(x_3 - 0.5)^2 + 10x_4 + 5x_5 + \varepsilon,
\end{equation}

\noindent where $\varepsilon \sim \mathcal{N}(0, 1)$ and the independent variables $x_j \sim \mathcal{U}(0, 1)$. Each penalized model produces a subset of variables, which is then used as input to five black-box algorithms: Random Forest, XGBoost, LightGBM, CatBoost, and H2O GBM. Each algorithm is trained three times, once for each subset of variables selected by Ridge, Lasso, and Elastic Net, resulting in 15 distinct hybrid models. Additionally, the same five algorithms are trained using all original variables without prior selection, resulting in a total of 23 distinct models. The models were trained using stratified cross-validation to ensure the robustness of the root mean square error (RMSE) metrics. The complete diagram can be seen at \ref{fig:simul_diagram}

\subsection{Simulation Results}\label{results:simulation}

To evaluate the robustness and generalizability of the proposed models, we conducted a simulation study with different sample sizes ($n = 200$, $500$, and $1000$) and total numbers of predictor variables ($p = 5$, $10$, $50$). When the number of potential predictors was greater than 5, i.e., above that provided in the model being inferred, the extra predictors were generated independently from uniform distributions and did not affect the data generation mechanism, being merely noise predictors. The goal of this was to understand whether the algorithms were able to discriminate predictors that were truly important to predict the response. Each scenario included regularized models, black-box models using all variables, and hybrid models combining variable selection through penalized regression with machine learning algorithms.

\begin{figure}[H]
    \centering
    \includegraphics[width=0.9\textwidth]{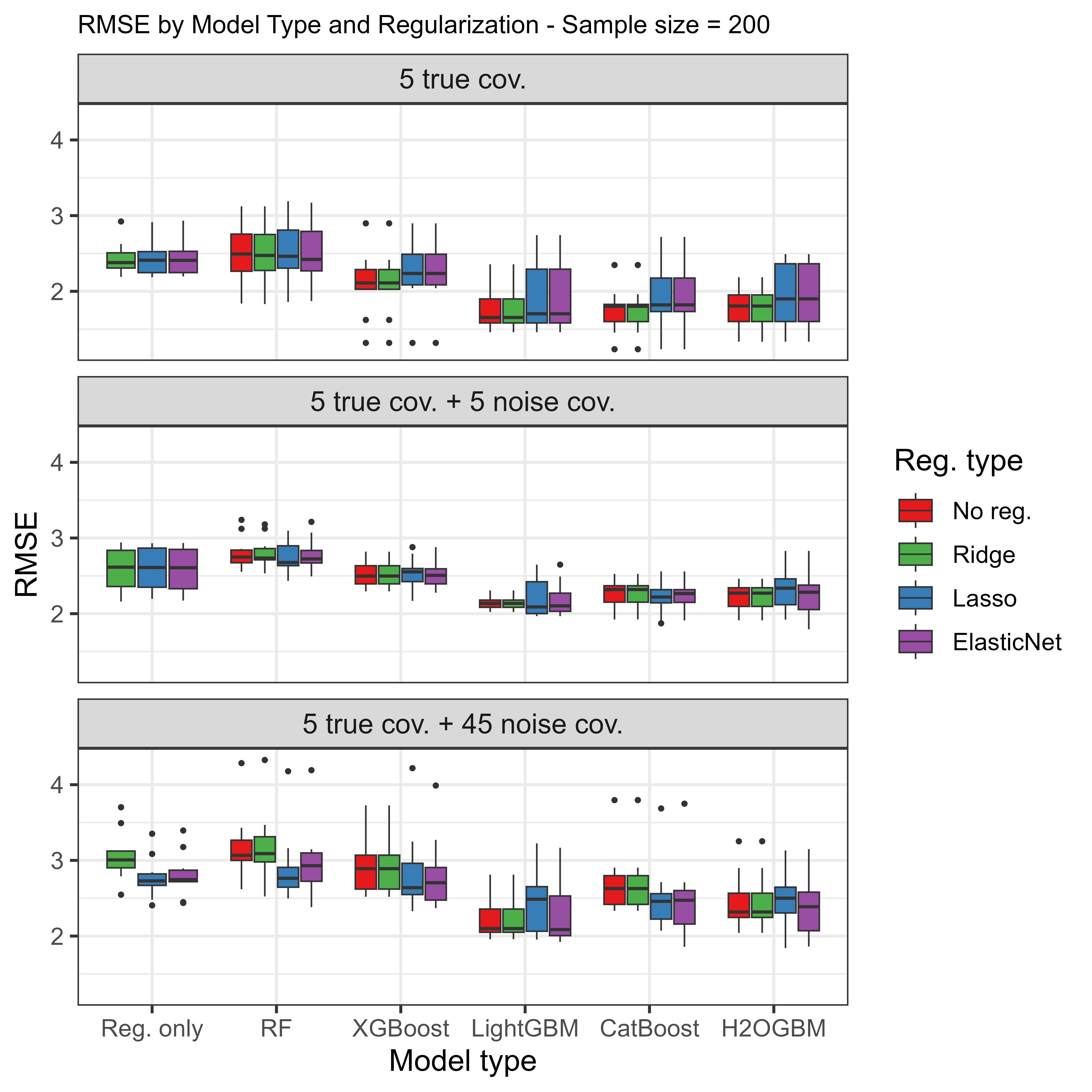}
    \caption{RMSE distribution by model – $n=200$ \textbf{(a)}}
    \label{fig:sim200}
\end{figure}

Figure ~\ref{fig:sim200} presents the results for $n=200$. It is evident that hybrid models based on CatBoost, LightGBM, and H2O GBM, combined with Elastic Net or Lasso, outperform, on average, both models trained on all variables and regularized linear models. The performance difference is particularly noticeable in low-dimensional scenarios ($p=5$), suggesting that variable selection is especially beneficial when fewer informative predictors are available.

\begin{figure}[H]
    \centering
    \includegraphics[width=0.9\textwidth]{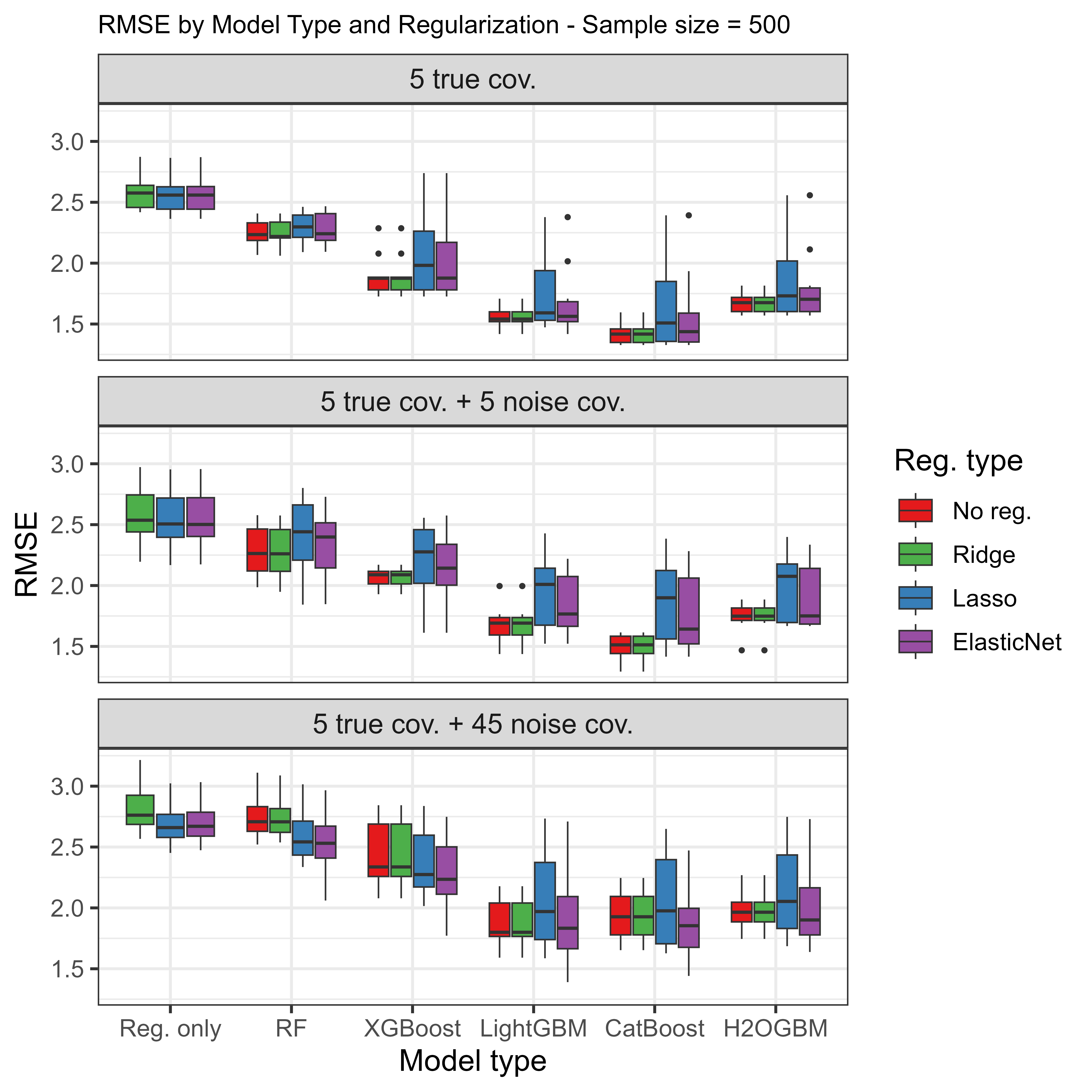}
    \caption{RMSE distribution by model – $n=500$ \textbf{(b)}}
    \label{fig:sim500}
\end{figure}

Figure~\ref{fig:sim500}, corresponding to the intermediate case ($n=500$), confirms the advantage of hybrid models, with \textit{CatBoost\_ElasticNet} and \textit{H2OGBM\_Lasso} achieving consistently low RMSE, even as dimensionality increases. Purely regularized models (\textit{Ridge}, \textit{Lasso}, \textit{Elastic Net}) begin to lose competitiveness as $p$ increases, exposing their limitations in capturing complex nonlinear interactions.

\begin{figure}[H]
    \centering
    \includegraphics[width=0.9\textwidth]{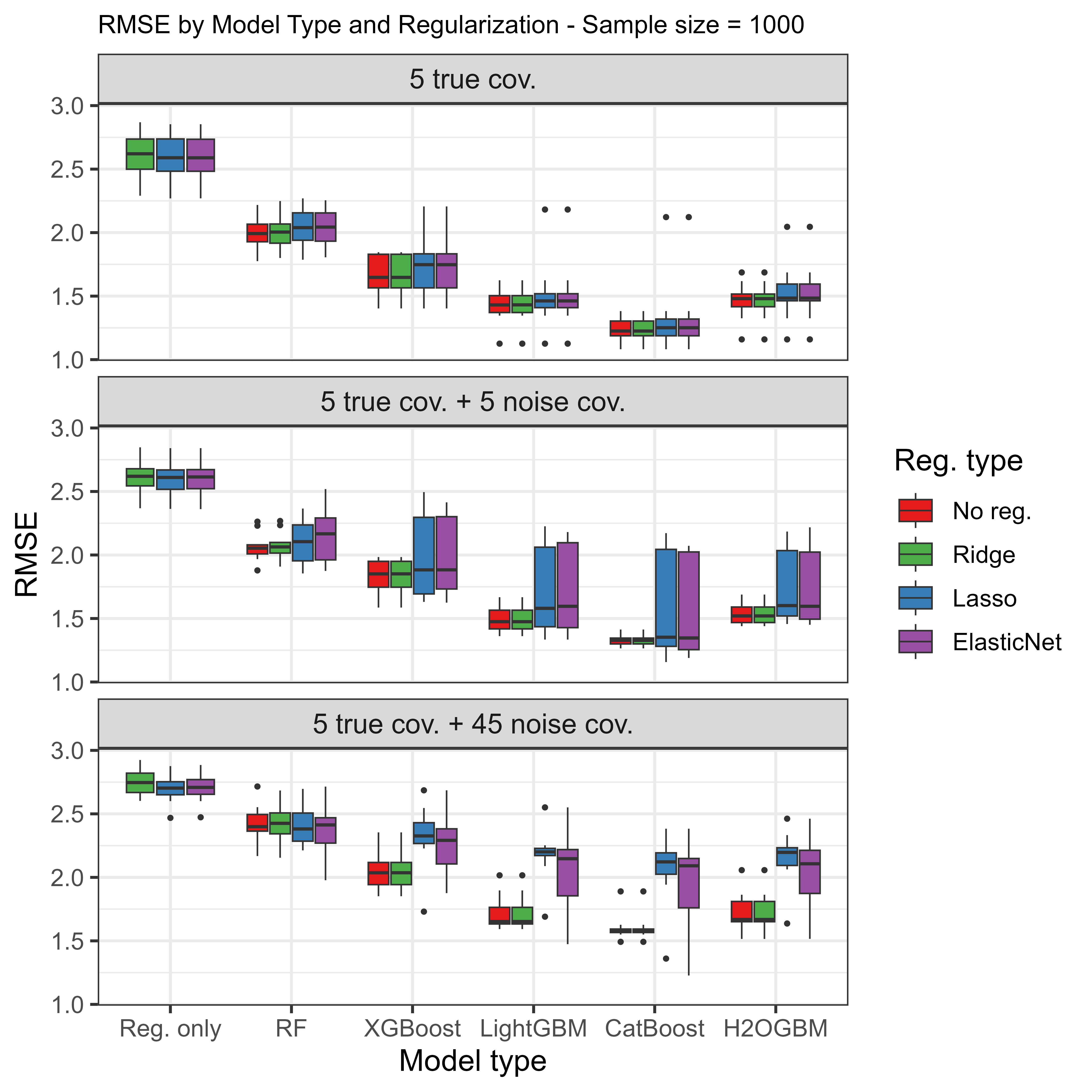}
    \caption{RMSE distribution by model – $n=1000$ \textbf{(c)}}
    \label{fig:sim1000}
\end{figure}

Figure ~\ref{fig:sim1000} shows the results for $n=1000$, the previous observation remains the same since the hybrid models achieve the lowest mean errors at almost all levels of complexity (variations of $p$). As the sample size increases, the black-box estimators stabilize, but the hybrid models still stand out for their superior balance between bias and variance, particularly in controlling the variability of the RMSE.

\begin{figure}[H]
    \centering
    \includegraphics[width=0.9\textwidth]{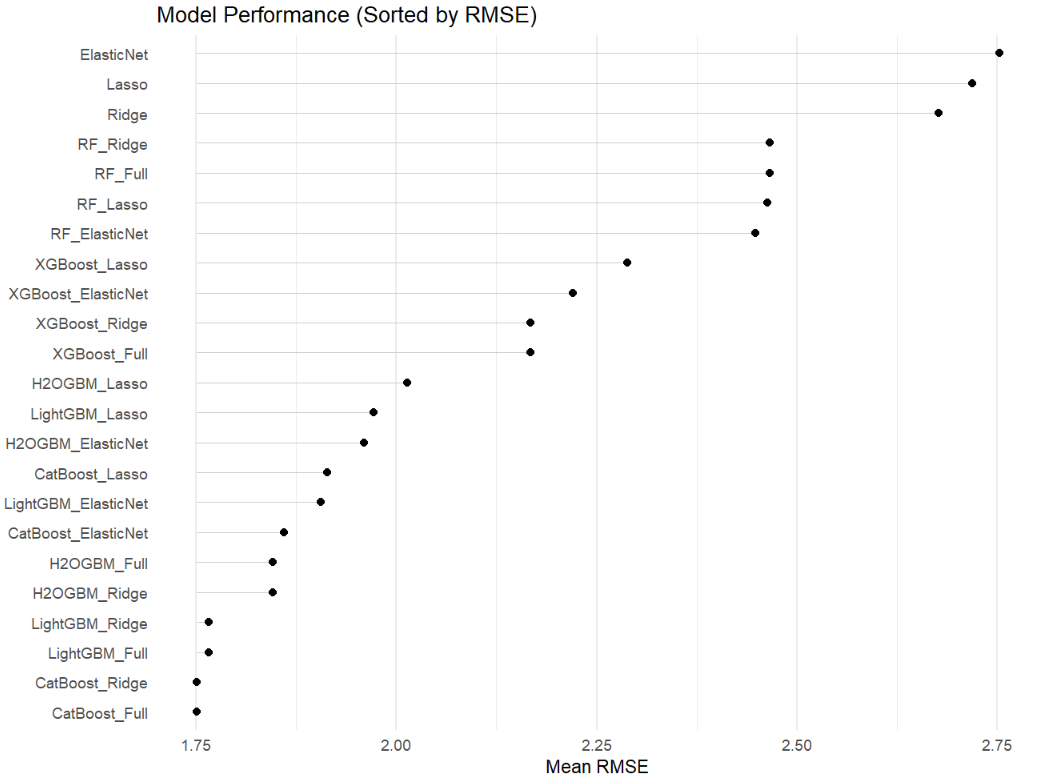}
  \caption{Comparison of predictive models ranked by their mean Root Mean Squared Error (RMSE). Lower values indicate better performance.}

    \label{fig:dotchart_rmse}
\end{figure}

The  \ref{fig:dotchart_rmse} summarizes the average RMSEs among the simulated models. The black-box models without consortium present a good predictive power but do not differ substantially in the image from their hybrid equivalents (Regularized + Black-Box Equivalent).

Simulation results indicate that combining penalized variable selection with Blackbox algorithms is an effective strategy for improving predictive performance. Hybrid models strike a balance between robustness and accuracy, making them particularly suitable for applications such as predicting travel insurance adoption, where the number of important and irrelevant variables can vary significantly.\\

\section{Conclusion}\label{conclusion}

Our results demonstrate that the performance of the methodologies is distinct from each other, with gains in specific aspects, whether precision or interpretability, and important empirical nuances of the sector were observed, depending on the focus, among the strategies considered.

The regularized linear models presented AUC values around 0.810 and F1 scores ranging from 0.73 to 0.74, indicating reliable predictive power, but with relatively inferior performance. It is worth noting that these models offer greater interpretability, which continues to be a critical advantage in contexts involving regulatory compliance, as is the case in the insurance sector. This is an environment where having extensive documentation and validation by experts is essential to launch any product in the insurance market.

In comparison, the pure black-box models, trained with the complete set of variables, obtained superior performance, in terms of F1 (which balances precision and recall). The LightGBM model (AUC = 0.902, F1 = 0.82) is recognized in the literature and by data bureaus in the insurance market for its great predictive power. As widely discussed, this model (and the others of this nonlinear profile) is suitable for production systems where real-time decision making and low error tolerance are essential. The hybrid models showed competitive performance, but slightly lower than their full-variable counterparts. The best among them was CatBoost with Lasso, achieving AUC = 0.8611 and F1 = 0.8082. Although this combination, in this application case and more generally in our simulations, slightly reduces the predictive power in the AUC, it offers better interpretability, reduced dimensionality and computationally faster inference times, important considerations in high-throughput environments, constrained service load times or when model transparency is required. In this scenario, hybrid models are a viable application option for the travel insurance sector and new work is encouraged in order to expand the possible scope of action of this methodology, which can be used in regression and classification problems.
\end{spacing}

\bibliographystyle{IEEEtran}

\vspace{0.60cm}

\section{Supplementary Materials}\label{AnexoI}

\begin{longtable}{|l|p{10cm}|}
\caption{Variable names and descriptions used in the predictive model} \\
\hline
\textbf{Variable name} & \textbf{Description} \\
\hline
\endfirsthead

\hline
\textbf{Variable name} & \textbf{Description} \\
\hline
\endhead

Age & Customer's age (years) \\
AnnualIncome & Reported annual gross income \\
FamilyMembers & Number of family members \\
GraduateOrNot & Indicates whether the customer holds a degree (Yes/No) \\
Employment.Type & Declared employment type (nominal) \\
ChronicDiseases & Number of reported chronic conditions \\
FrequentFlyer & Participates in a frequent flyer program (Yes/No) \\
EverTravelledAbroad & Has previously traveled abroad (Yes/No) \\
TravelInsurance & Target variable: purchased travel insurance (Yes = 1, No = 0) \\
IncomePerCapita & Annual income divided by the number of family members \\
HighIncome & Income above the 75th percentile \\
AgeNormalized & Standardized age (z-score) \\
HighChronicDiseases & Indicator for number of chronic diseases above the median \\
TravelFrequency & Indicator based on either frequent flyer status or prior travel abroad \\
PrivateEmployment & Binary indicator for private/self-employed sector \\
LowDependence & Indicator for small family size (3 or fewer members) \\
IncomeByAge & Income-to-age ratio \\
AgeGroup & Categorized age groups: young, adult-y, H-age, senior \\
HighIncomeTraveler & High-income individual with travel history \\
HighIncome90 & Income above the 90th percentile \\
IncomePerCapitaNorm & Standardized income per capita (z-score) \\
ExperiencedTraveler & Indicator for prior international travel \\
LargeFamily & Family size above the 75th percentile \\
ChronicByAge & Ratio of chronic diseases to age \\
InsuranceScore & Composite score based on income, chronic diseases, travel frequency, and experience \\
FinancialDependence & Log-transformed family size to income ratio \\
TravelScore & Discrete score based on travel frequency and experience \\
WorkExperience & Estimated years of professional experience based on age and graduation status \\
StableJob & Binary indicator for government sector employment \\
AdjustedTravelIncome & Income adjusted by travel frequency (normalized) \\
RiskScore & Composite score based on chronic diseases, age, and travel frequency \\
RiskScoreNorm & Standardized risk score (z-score) \\
ClusterScore & Cluster label based on k-means grouping of income, age, and employment \\
ClusterInsuranceRate & Average insurance uptake rate per cluster \\
MovingAvgInsurance & Moving average of insurance uptake by age group \\
\hline
\end{longtable}

\begin{longtable}{|l|p{6cm}|p{4.5cm}|}
\caption{Variable names, descriptions, and classification used in the predictive model} \\
\hline
\textbf{Variable name} & \textbf{Description} & \textbf{Origin and Type} \\
\hline
\endfirsthead

\hline
\textbf{Variable name} & \textbf{Description} & \textbf{Origin and Type} \\
\hline
\endhead

Age & Customer's age (in years) & Original – Numeric \\
AnnualIncome & Declared gross annual income & Original – Numeric \\
FamilyMembers & Total family members & Original – Numeric \\
GraduateOrNot & Has a college degree (Yes/No) & Original – Binary \\
Employment.Type & Declared job type & Original – Categorical \\
ChronicDiseases & Number of chronic conditions & Original – Numeric \\
FrequentFlyer & Enrolled in mileage program & Original – Binary \\
EverTravelledAbroad & Traveled abroad before & Original – Binary \\
TravelInsurance & Target: insurance purchased & Original – Binary \\
IncomePerCapita & Income per family member & Derived – Numeric \\
HighIncome & Income  75th percentile & Derived – Binary \\
AgeNormalized & Standardized age (z-score) & Derived – Numeric \\
HighChronicDiseases & Above-median chronic conditions & Derived – Binary \\
TravelFrequency & Flyer or past travel flag & Derived – Binary \\
PrivateEmployment & Private/self-employed flag & Derived – Binary \\
LowDependence & Family size $\leq 3$ & Derived – Binary \\
IncomeByAge & Income divided by age & Derived – Numeric \\
AgeGroup & Age grouped by range & Derived – Categorical \\
HighIncomeTraveler & High income and traveler & Derived – Binary \\
HighIncome90 & Income > 90th percentile & Derived – Binary \\
IncomePerCapitaNorm & Normalized per capita income & Derived – Numeric \\
ExperiencedTraveler & Has traveled abroad & Derived – Binary \\
LargeFamily & Family size > 75th percentile & Derived – Binary \\
ChronicByAge & Chronic cond. per age & Derived – Numeric \\
InsuranceScore & Composite insurance score & Derived – Numeric \\
FinancialDependence & log(Family / Income) & Derived – Numeric \\
TravelScore & Score from travel activity & Derived – Discrete \\
WorkExperience & Estimated work years & Derived – Numeric \\
StableJob & Government job flag & Derived – Binary \\
AdjustedTravelIncome & Travel-based adj. income & Derived – Numeric \\
RiskScore & Composite risk score & Derived – Numeric \\
RiskScoreNorm & Standardized risk score & Derived – Numeric \\
ClusterScore & Cluster group label (k-means) & Derived – Categorical \\
ClusterInsuranceRate & Mean insurance per cluster & Derived – Numeric \\
MovingAvgInsurance & Avg. insurance by age group & Derived – Numeric \\
\hline
\end{longtable}

\end{document}